# Polynomial-Time Approximation Schemes for Knapsack and Related Counting Problems using Branching Programs

Parikshit Gopalan[*1], Adam Klivans[†2], and Raghu Meka[‡2]

[1]Microsoft Research, Silicon Valley.
[2]Department of Computer Science, UT Austin.

**Abstract**

We give a deterministic, polynomial-time algorithm for approximately counting the number of $\{0,1\}$-solutions to any instance of the knapsack problem. On an instance of length $n$ with total weight $W$ and accuracy parameter $\varepsilon$, our algorithm produces a $(1+\varepsilon)$-multiplicative approximation in time $\mathsf{poly}(n, \log W, 1/\varepsilon)$. We also give algorithms with identical guarantees for general integer knapsack, the multidimensional knapsack problem (with a constant number of constraints) and for contingency tables (with a constant number of rows). Previously, only randomized approximation schemes were known for these problems due to work by Morris and Sinclair and work by Dyer.

Our algorithms work by constructing small-width, read-once branching programs for approximating the underlying solution space under a carefully chosen distribution. As a byproduct of this approach, we obtain new query algorithms for learning functions of $k$ halfspaces with respect to the uniform distribution on $\{0,1\}^n$. The running time of our algorithm is polynomial in the accuracy parameter $\varepsilon$. Previously even for the case of $k=2$, only algorithms with an exponential dependence on $\varepsilon$ were known.

[*]Mail: parik@microsoft.com
[†]Mail: klivans@cs.utexas.edu. Work done while visiting Microsoft Research, Silicon Valley.
[‡]Mail: raghu@cs.utexas.edu. Work done as an intern at Microsoft Research, Silicon Valley.

# 1  Introduction

In this paper we give the first deterministic, polynomial-time approximation schemes for several well-studied #P-hard counting problems such as knapsack and multidimensional knapsack. There are many celebrated, randomized polynomial-time algorithms for approximately counting #P-hard problems (for example, approximating the permanent [JSV04]). There are far fewer examples, however, of *deterministic* approximation algorithms for #P-hard problems. A few notable examples can be found in [Wei06], [BGK+07], [HKM+09].

The knapsack counting problem (#KNAP) is defined as follows: given a non-negative vector $a \in \mathbb{Z}_+^n$ and non-negative $b \in \mathbb{Z}_+$, count the size of the set $\mathsf{KNAP}(a,b) = \{x \in \{0,1\}^n : \sum_i a_i x_i \leq b\}$. It is well-known that the #KNAP problem is #P-hard, and much attention has been given to the problem of approximately counting the size of $\mathsf{KNAP}(a,b)$. More specifically, given an error parameter $\varepsilon$, we are interested in finding a value $p$ such that $|\mathsf{KNAP}(a,b)| \leq p \leq (1+\varepsilon)|\mathsf{KNAP}(a,b)|$ in time polynomial in $n$ and $1/\varepsilon$ (such a value $p$ is often referred to as an $\varepsilon$ *relative-error* approximation or $\varepsilon$-approximation for short).

Dyer et al. [DFK+93] were the first to study the problem of approximately solving #KNAP and gave a sub-exponential time algorithm for the problem. Morris and Sinclair [MS04] were the first to give a polynomial-time, randomized approximation scheme (FPRAS) for #KNAP, they use a rapidly mixing Markov chain to sample randomly from the solution space. Subsequently, Dyer [Dye03] gave a simpler FPRAS based on dynamic programming for #KNAP. In this work, we give the first *deterministic* polynomial-time approximate counting algorithm for #KNAP[1]:

**Theorem 1.1** (determnistic counting for knapsack). *Given a knapsack instance* $(a,b) \in \mathbb{Z}_+^{n \times 1}$ *with weight* $W = \sum_i a_i + b$ *and* $\varepsilon > 0$, *there is a deterministic* $O(n^3 \log(W) \log(n/\varepsilon)/\varepsilon)$ *time algorithm that computes an $\varepsilon$-relative error approximation for* $|\mathsf{KNAP}(a,b)|$.

Our algorithm is simple and yields a fast method for generating a uniformly random element of $\mathsf{KNAP}(a,b)$. The algorithm is inspired by a recent work due to Meka and Zuckerman [MZ10] on monotone branching programs in the context of building pseudorandom generators for halfspaces. Further, we show how to extend our algorithm to work with respect to a broad class of natural non-uniform distributions on $\{0,1\}^n$ including all symmetric and product distributions. To the best of our knowledge, no efficient algorithms (randomized or otherwise) for counting with respect to these natural distributions were known previously (see Section 1.1 for more details).

Morris and Sinclair [MS04] and Dyer [Dye03] also gave an FPRAS for counting the number of solutions to the *multidimensional knapsack problem* for a constant number of constraints. In this problem, we are given $k$ knapsack instances $(a_1, b_1), \ldots, (a_k, b_k) \in \mathbb{Z}_+^{n \times 1}$, $\varepsilon > 0$, and the goal is to compute the number of solutions satisfying all constraints; i.e., compute $|\mathsf{KNAP}(a_1, b_1) \cap \mathsf{KNAP}(a_2, b_2) \cap \cdots \cap \mathsf{KNAP}(a_k, b_k)|$. We obtain a deterministic algorithm for this problem that runs in polynomial time for $k = O(1)$:

**Theorem 1.2** (multi-dimensional knapsack). *Given knapsack instances* $\mathsf{KNAP}(a_1, b_1), \ldots, \mathsf{KNAP}(a_k, b_k)$ *of total weight at most* $W$, *there is a deterministic* $O((n/\varepsilon)^{O(k^2)} \cdot \log W)$ *algorithm that computes an $\varepsilon$-relative error approximation to the number of solutions satisfying all the knapsack constraints.*

Our solution has two components: we generalize our counting algorithm for a single knapsack constraint to work with respect to non-uniform distributions representable by small-width branching programs. We then use Dyer's elegant rounding results to reduce multidimensional knapsack counting to counting a single knapsack under such distributions.

---

[1]Throughout this paper we assume integer addition to be of unit cost.

Our techniques also apply to the problem of counting the solutions of integer-valued knapsack instances. Here the goal is to estimate the size of the set of solutions $\mathsf{KNAP}(a,b,u) = \{x : \sum_{i \leq n} a_i x_i \leq b, \ 0 \leq x_i \leq u_i\}$. Note that the range sizes $u_1, \ldots, u_n$ could be exponential in $n$. Dyer [Dye03] gave an FPRAS for the integer-valued case as well. We obtain a FPTAS for the problem.

**Theorem 1.3** (integer knapsack). *Given a knapsack instance $\mathsf{KNAP}(a,b,u)$ with weight $W = \sum_i a_i u_i + b$, $U = \max_i u_i$ and $\varepsilon > 0$, there is a deterministic $O(n^5 (\log U)^2 (\log W)/\varepsilon^2)$ algorithm that computes an $\varepsilon$-relative error approximation for $|\mathsf{KNAP}(a,b,u)|$.*

We also obtain similar results for counting the number of integer-valued contingency tables. Given row sums $r = (r_1, \ldots, r_m) \in \mathbb{Z}_+^m$ and column sums $c = (c_1, \ldots, c_n) \in \mathbb{Z}_+^n$, let $CT(r,c) \subseteq \mathbb{Z}_+^{m \times n}$ denote the set of integer-valued contingency tables with row and column sums given by $r, c$:

$$CT(r,c) = \{ X \in \mathbb{Z}_+^{m \times n} : \sum_j X_{ij} = r_i, i \in [m], \ \sum_i X_{ij} = c_j, j \in [n] \}.$$

Note that as in the case of knapsack, the magnitude of the row and column sums could be exponential in $n$. Dyer [Dye03] gave an FPRAS for counting solutions to contingency tables (with a constant number of rows) based on dynamic programming. We give a FPTAS for this problem:

**Theorem 1.4** (contingency tables with few rows). *Given row sums $r = (r_1, \ldots, r_m) \in \mathbb{Z}_+^m$ and column sums $c = (c_1, \ldots, c_n) \in \mathbb{Z}_+^n$ with $R = \max_i r_i$ and $\varepsilon > 0$, there is a deterministic $(n^{O(m)} (\log R)/\varepsilon)^m$ algorithm that computes a $\varepsilon$-relative error approximation for $|CT(r,c)|$.*

All our counting results also give fast sampling algorithms. After a pre-processing phase, each new random sample can be generated in near-linear time, which improves considerably on previous sampling algorithms.

Finally, we can use ideas motivated by our algorithm for counting knapsack solutions to learn functions of halfspaces with membership queries with respect to the uniform distribution on $\{0,1\}^n$:

**Theorem 1.5.** *The concept class of arbitrary Boolean functions of k halfspaces can be PAC learned with membership queries under the uniform distribution $\{0,1\}^n$ to accuracy $\varepsilon$ in time $(n/\varepsilon)^{O(k)}$.*

Previous algorithms [KOS04] ran in time $n^{O(k^2/\varepsilon^2)}$ (without queries) or in time $\mathsf{poly}(n^{2^k}, W^{2^k}, 1/\varepsilon)$ (with queries) where $W$ is a bound on the weight of all halfspaces (which could be exponential in $n$). Thus, even for the special case of learning the intersection of two halfspaces, known algorithms had either an exponential dependence on $1/\varepsilon$ or a polynomial dependence on $W$. Our algorithm is similar to Angluin's algorithm for learning finite automata [Ang87] (essentially we reconstruct the underlying approximating branching program). The analysis however, is quite different, since we learn functions that are not (exactly) computable by small-width ROBPs.

## 1.1 Outline of our Algorithms.

**Approximation by Branching Programs.** All of our results revolve around the ability of read-once branching programs (ROBPs) to approximate various classes of Boolean functions. Informally, ROBP of width $W$ is a labeled, layered directed graph with at most $W$ vertices per layer that induces a Boolean function in the obvious manner: at layer $i$ we read the $i$'th bit of input, follow the appropriate transition, and output the label of the final vertex reached (see Definition 2.1).

It is easy to see that a knapsack constraint of weight $W$ (recall $W$ may be exponential in $n$) can be computed exactly by a width-$W$ ROBP which keeps track of partial sums. Meka and Zuckerman [MZ10] in their work on pseudorandom generators for halfspaces[2] proved the existence of a small-width ROBP that approximates the solutions to a single knapsack constraint with small additive (as opposed to multiplicative) error. To give an algorithm for approximately counting knapsack solutions, we show how to *explicitly* construct a small-width ROBP whose set of accepting strings is a multiplicative approximation to the set of strings satisfying the knapsack constraint. Our construction proceeds by sparsifying each layer in the exact ROBP for knapsack by retaining only a few carefully chosen representative partial sums. This uses the insight from [MZ10] that in the exact branching program for halfspaces, there is a natural ordering on the vertices in each layer induced by the partial sums.

Building on these ideas, we give a query algorithm that can learn an unknown Boolean function under the uniform distribution as long as it is approximated in a certain sense by a small-width ROBP. Previous learning algorithms (e.g., Angluin's algorithm for learning finite automata) required the function to be exactly computable by a small-width ROBP. Our notion of approximation is somewhat subtle: it is stronger than being approximated by an ROBP under the uniform distribution, but weak enough that any function of few halfspaces has such approximations.

**Small-Space Sources.** Extending our knapsack algorithm to multiple knapsack constraints is not immediate. The main obstacle is that the natural ROBP for the intersection of knapsack constraints which keeps track of all partial sums does not have a total ordering on its vertices, and our knapsack algorithm crucially uses such a total ordering. One can still construct a small width ROBP that additively approximates the number of solutions to multidimensional knapsack (as in [GOWZ10]), but even the existence of a small-width multiplicative approximation is unclear.

To circumvent this issue, we first generalize our algorithms to counting knapsack solutions with respect to *small-space sources* which were introduced by Kamp et al. [KRVZ06] in the context of randomness extraction. Informally, these are families of distributions on $\{0,1\}^n$ that can be generated by small-width branching programs (see Section 2 for the formal definition). We show the following result for deterministically counting knapsack solutions under small space sources:

**Theorem 1.6** (counting under small-space sources). *Fix a knapsack instance of total weight $W$ and error parameter $\varepsilon > 0$. Let $\mu$ be a distribution on $\{0,1\}^n$ with an explicitly given small space generator of width at most $S$ and define $\mu(\mathsf{KNAP}(a,b))$ as $\Pr_{x \sim \mu}[x \in \mathsf{KNAP}(a,b)]$. Then there is a deterministic algorithm that runs in time $O(n^3 S(S + \log W) \log(n/\varepsilon)/\varepsilon)$ and computes an $\varepsilon$-relative error approximation to $\mu(\mathsf{KNAP}(a,b))$.*

Next, we use an elegant result of Dyer [Dye03], which given a instance of multidimensional knapsack, constructs a small space source under which the set of solutions is polynomially dense. It is easy to get an additive approximation for multidimensional knapsack using Theorem 1.6. Dyer's result lets us transform a low additive error guarantee into a multiplicative error guarantee and prove Theorem 1.2.

Small-space distributions include several natural distributions such as all symmetric distributions and product distributions. Thus as a corollary to Theorem 1.6, we obtain FPTASes for several interesting variants of knapsack for which no polynomial time algorithms – even randomized – were known to the best of our knowledge. For instance we show:

**Corollary 1.7.** *Given a knapsack instance $(a,b) \in \mathbb{Z}_+^{n \times 1}$ of total weight $W = \sum_i a_i + b$, $\varepsilon > 0$ and $r \in [n]$ we can in deterministic time $O(n^3 r(r + \log W)/\varepsilon)$ compute an $\varepsilon$-relative error approximation for the number of solutions to the knapsack instance of Hamming weight exactly $r$.*

---

[2]Halfspaces are equivalent to the characteristic functions of knapsack instances.

## 1.2 Related Work

Very recently, Stefankovic, Vempala, and Vigoda [SVV10] gave a deterministic FPTAS for the knapsack problem with a run-time of $O(n^3 \varepsilon^{-1} \log(n/\varepsilon))$. Their algorithm is based on dynamic programming. Our work was obtained independently of theirs.

As mentioned earlier, Morris and Sinclair [MS04] and Dyer [Dye03] were the first to give an FPRAS for the knapsack problem, with Dyer's more efficient algorithm taking time $O(n^{2.5}\sqrt{\log(n/\varepsilon)} + n^2 \varepsilon^{-2})$. Dyer also gives a deterministic, polynomial-time algorithm that achieves a $\sqrt{n}$ factor approximation for counting knapsack solutions.

The problem of approximately counting knapsack solutions is equivalent to the problem of approximately counting the fraction of assignments that satisfy a linear threshold function or *halfspace*. Servedio [Ser07] gave a deterministic algorithm for solving the latter problem to within an *additive* $\varepsilon$ in time exponential in $1/\varepsilon^2$. Recently, Diakonikolas et al. [DGJ$^+$09] gave a pseudorandom generator for halfspaces with seed-length $\tilde{O}(\log n/\varepsilon^2)$ and Meka and Zuckerman [MZ10] gave a pseudorandom generator for halfspaces with seed-length $O(\log n \log(1/\varepsilon))$ (enumerating over all seeds results in a deterministic, additive error approximation).

Many researchers in computational learning theory have studied the problem of learning functions computable by read-once branching programs (for a discussion see Bshouty et al. [BTW98]). Positive results were known only for restricted classes of ROBPs, such as width-2 ROBPs [EKR95, BTW98] (these algorithms use queries and succeed in the distribution-free model of learning) and do not apply in our setting. Our algorithms learn concept classes that are closely approximated by small-width ROBPs (with respect to the uniform distribution on $\{0,1\}^n$).

## 2 Preliminaries

### 2.1 Read-Once Branching Programs

**Definition 2.1** (ROBP). *An $(S,T)$-branching program $M$ is a layered multi-graph with a layer for each $0 \leq i \leq T$ and at most $S$ vertices (states) in each layer. The first layer has a single vertex $v_0$ and each vertex in the last layer is labeled with $0$ (rejecting) or $1$ (accepting). For $0 \leq i \leq T$, a vertex $v$ in layer $i$ has two outgoing edges labeled $0,1$ and ending at vertices in layer $i+1$.*

Note that by definition, an $(S,T)$-branching program is read-once. We also use the following notation. Let $M$ be an $(S,T)$-branching program and $v$ a vertex in layer $i$ of $M$.

1. For a string $z$, $M(v,z)$ denotes the state reached by starting from $v$ and following edges labeled with $z$.

2. For $z \in \{0,1\}^n$, let $M(z) = 1$ if $M(v_0, z)$ is an accepting state, and $M(z) = 0$ otherwise.

3. $A_M(v) = \{z : M(v,z)$ is accepting in $M\}$ and $P_M(v)$ is the probability that $M(v,z)$ is an accepting state for $z$ chosen uniformly at random.

4. $L(M,i)$ denotes the vertices in layer $i$ of $M$.

5. For a set $U$, $x \in_u U$ denotes a uniformly random element of $U$.

### 2.2 Small-Space Sources

Small-space sources were introduced by Kamp et al. [KRVZ06] in their work on randomness extractors, as a generalization of many commonly studied distributions such as bit-fixing sources, Markov-chain sources.

**Definition 2.2** (small-space sources, Kamp et al.)**.** *A width $w$ small-space source is described by a $(w, n)$-branching program $D$ with an additional probability distribution $p_v$ on the outgoing edges associated with vertices $v \in D$. Samples from the source are generated by taking a random walk on $D$ according to $p_v$'s and outputting the labels of the edges traversed.*

We will often abuse notation and denote the distribution generated by a small-space source and the source itself by $D$. Also, we will assume that the distribution $D$ is given to us explicitly as a small-space source. Several natural distributions such as all symmetric distributions and product distributions can be generated by a small-space source. The following claims are straightforward:

**Claim 2.3.** *Given a ROBP $M$ of width at most $W$ and a small-space source $D$ of width at most $S$, $\Pr_{x \leftarrow D}[M(x) = 1]$ can be computed exactly via dynamic programming in time $O(n \cdot S \cdot W)$.*

**Claim 2.4.** *Given a $(W, n)$-ROBP $M$, the uniform distribution over $M$'s accepting inputs, $\{x : M(x) = 1\}$ is a width $W$ small-space source.*

## 3 A FPTAS for Counting Knapsack Solutions

As described in the introduction, we construct a small-width branching program that approximates the feasible solutions to the given knapsack instance. We start with the exact branching program for knapsack which has width $W$, and where each state in layer $j$ corresponds to a possible value of the partial sum $v_j = \sum_{i \leq j} a_i x_i$. We will approximate this program with a small width branching program whose state space is a carefully chosen subset of the original state space. We then count the number of accepting solutions to the constructed small-width branching program exactly via dynamic programming.

Our goal is to partition the states in layer $i$ into intervals $I_1 = \{v_1 = 0, \ldots, v_2 - 1\}, I_2 = \{v_2, \ldots, v_3 - 1\}, I_t = \{v_t, \ldots, v_{t+1} = W\}$ and have only one state for each interval. The intervals should be such that the number of accepting suffixes for all the partial sums in an interval is roughly the same. We then rearrange the incoming edges from layer $i - 1$ appropriately. We refer to this process as *rounding* layer $i$. A problem with this approach is that counting the number of suffixes which accept from a given partial sum is another instance of knapsack.

We handle this by building the small width branching program backward starting from the last layer. When we round the layer $i$, the layers $i + 1, \ldots, n$ have already been rounded. Thus given a partial sum in layer $i$, we know the number of accepting suffixes in our branching program exactly and use these counts to partition layer $i$. We then show by induction that the resulting branching program gives a good approximation to the set of feasible knapsack solutions.

We now give a formal description of this process.

### 3.1 Constructing an Approximating Branching Program

Let $M$ denote the exact branching program for $\sum_{i \leq n} a_i x_i \leq b$, which consists of $n + 1$ layers numbered from 0 to $n$. We denote the set of states in layer $i$ by $L(M, i)$. Layer 0 has a single start state $s$. For $i \leq n$, $L(M, i)$ has a state for each partial sum $\sum_{j \leq i} a_j x_j$. Given a vertex $v$ in layer $i - 1$ and $x_i \in \{0, 1\}$, the $x_i$'th neighbor of $v$ $M(v, x_i) = v + a_i x_i$.

We construct a series of branching programs $M^n = M, M^{n-1}, \ldots, M^0$. We obtain $M^i$ from $M^{i+1}$ by rounding the states in $L(M^{i+1}, i+1)$. More precisely, we set $L(M^i, i+1) = \{v_1, \ldots, v_\ell\} \subseteq L(M^{i+1}, i+1)$ where the $v_j$s are defined as follows: Let $v_1 = 0$. Given $v_j$, let

$$v_{j+1} = \min \ v \text{ such that } v > v_j \text{ and } 0 < P_{M^{i+1}}(v) \leq P_{M^{i+1}}(v_j)/(1 + \varepsilon) \tag{3.1}$$

Intuitively, state $v_j$ represents the interval $I_j = \{v_j, \ldots, v_{j+1}-1\}$. When the acceptance probability drops by a factor of $(1+\varepsilon)$, we start a new interval. Since $P_{M^i}(v_1) \leq 1$ and $P_{M^i}(v_{\ell-1}) \geq 2^{-n}$, we have $\ell \leq O(\frac{n}{\varepsilon})$. Next we redirect the edges from level $i$ to level $i+1$. If there is an edge labeled $z$ entering a vertex $v \in I_j$, then we redirect the edge to vertex $v_j$. Note that rounding layer $n$ to get $M^{n-1}$ is trivial, we keep just one accept state and one reject state, corresponding to partial sums of $0$ and $b+1$ respectively.

Our branching programs have the following monotonicity property which is easily verified by induction. We omit the proof.

**Lemma 3.1.** *Let $v, v' \in L(M^i, j)$ and $v \leq v'$. For any suffix $z$, $M^i(v, z) \leq M^i(v', z)$. Hence $P_{M^i}(v) \geq P_{M^i}(v')$.*

This property allows us to construct $M^i$ from $M^{i+1}$ efficiently. The key idea is that in Equation A.1, due to the ordering of the probabilities $P_M()$, we can find $v_j$ by binary search as opposed to sequential search, reducing the running time to $O(\log W)$ as opposed to $O(W)$.

**Lemma 3.2.** *Each vertex $v_j \in L(M^i, i+1)$ can be computed in time $O(\log(n/\varepsilon) \log W)$.*

*Proof.* The prove is by induction: we maintain the invariant that for every $i$, we have the vertices $v_j$ of $L(M^i, i+1)$ stored in a binary tree and also know their acceptance probabilities $P_{M^i}()$.

Suppose we have the above setup for $l > i$ and have computed $v_1, \ldots, v_j \in L(M^i, i+1)$. Recall that $v_{j+1}$ is the smallest value of $v > v_j$ satisfying $P_{M^{i+1}}(v) < P_{M^{i+1}}(v_j)/(1+\varepsilon)$. Given a vertex $v \in L(M^{i+1}, i+1)$, if $v_b = M^{i+1}(v, b)$ for $b \in \{0, 1\}$, then $P_{M^{i+1}}(v) = (P_{M^{i+1}}(v_0) + P_{M^{i+1}}(v_1))/2$. So $P_{M^{i+1}}(v)$ can be computed in time $O(\log(n/\varepsilon))$ using the values of $P_{M^{i+1}}(w)$ stored in a binary tree for $w \in L(m^{i+1}, i+2)$.

Lemma 3.1 shows that $P_m(v)$ decreases as $v$ increases. So we can do binary search on $P_{M^{i+1}}(v)$. Since $v \in \{0, \ldots, W\}$, this will require $O(\log(W))$ computations of $P_{M^{i+1}}(v)$. Once we have computed $L(M^i, i+1)$ we store these vertices and their probabilities of acceptance in a binary tree. □

Thus, we can construct $M^0$ from $M$ in time $O(n^2 \log(W) \log(n/\varepsilon)/\varepsilon)$. We now address the approximation guarantee. We start by showing that the set of strings accepted grows as we proceed from $M$ to $M^0$.

**Lemma 3.3.** *For $v \in M^i$, we have $A_{M^{i+1}}(v) \subseteq A_{M^i}(v)$. Thus $P_{M^i}(v) \geq P_M(v)$.*

*Proof.* Note that the claim is only interesting for $v \in L(M^i, j)$ where $j \leq i$, since for $j \geq i+1$, both $M^i$ and $M^{i+1}$ make identical transitions from $v$, and so $A_{M^i}(v) = A_{M^{i+1}}(v)$.

Let $j = i$. Let $v_b = M^i(v, b)$ for $b \in \{0, 1\}$. Since $M^i$ is obtained from $M^{i+1}$ by rounding layer $i+1$, there are vertices $v'_b = M^{i+1}(v, b)$ for $b \in \{0, 1\}$ in $L(M^{i+1}, i+1)$ such that $v'_b \geq v_b$, hence $A_{M^{i+1}}(M^{i+1}(v, b)) = A_{M^{i+1}}(v'_b) \subseteq A_{M^{i+1}}(v_b) = A_{M^i}(M^i(v, b))$. Thus the set of accepting suffixes can only increase for either choice of $b$, and the claim is proved.

The claim for $j < i$ follows since $M^i$ and $M^{i+1}$ are identical up to layer $i$, and for every $v \in L(M^i, i)$ we have $A_{M^{i+1}}(v) \subseteq A_{M^i}(v)$. □

Next we show that the set of accepting strings does not grow by too much.

**Lemma 3.4.** *For any vertex $v \in M^i$, we have $P_{M^i}(v) \leq P_M(v)(1+\varepsilon)^{n-i}$.*

*Proof.* It is sufficient to show that for every $i < n$ and $v \in M^i$, $P_{M^i}(v) \leq P_{M^{i+1}}(v)(1+\varepsilon)$. Let $v \in L(M^i, j)$. The above is trivial when $j \geq i+1$, since $A_{M^i}(v) = A_{M^{i+1}}(v)$ for such vertices $v$. Indeed, it suffices to consider the case when $j = i$, since for $j < i$, $M^i$ and $M$ are identical up to layer $i$. Hence we can express both $P_{M^{i+1}}(v)$ and $P_{M^i}(v)$ as the same convex combination of acceptance probabilities of vertices in layer $i$.

Let $j = i$. Fix a vertex $v \in L(M^i, i)$. Let $v_b = M^i(v, b)$ for $b \in \{0, 1\}$ be the vertices reached in $L(M^i, i+1)$. Since $M^i$ is obtained from $M^{i+1}$ by rounding the $i^{th}$ layer, there are vertices $v'_b = M^{i+1}(v, b)$ for $b \in \{0, 1\}$ in $L(M^{i+1}, i+1)$ such that $P_{M^{i+1}}(v_b) \leq (1+\varepsilon) P_{M^{i+1}}(v'_b)$. Thus

$$P_{M^i}(v) = \frac{1}{2}(P_{M^i}(v_0) + P_{M^i}(v_1)) \leq (1+\varepsilon) \frac{P_{M^{i+1}}(v'_0) + P_{M^{i+1}}(v'_1)}{2} = (1+\varepsilon) P_{M^{i+1}}(v).$$

$\square$

We can now finish the proof of Theorem 1.1.

*Proof of Theorem 1.1.* We set $\varepsilon = \Omega(\frac{\delta}{n})$ so that $(1+\varepsilon)^n \leq (1+\delta)$. Using Lemma 3.2, we can construct $M^0$ and compute $P_{M^0}(s)$ where $s$ is the start state in the desired time bound. Applying Lemma 3.3 and Lemma 3.4 we get $P_M(s) \leq P_{M^0}(s) \leq (1+\delta) P_M(s)$. The number of knapsack solutions is precisely $2^n P_M(s)$. Hence we output $2^n P_{M^0}(s)$. $\square$

We note that our algorithm also gives an efficient sampling scheme, since sampling from the set of accepting strings of a small-width branching program is easy.

**Theorem 3.5.** *There is a randomized algorithm which produces a uniformly random string from the set of solutions to a knapsack instance $\mathsf{KNAP}(a, b)$. The algorithm takes a processing time of $O(n^3 \log(W) \log n)$ and then produces a uniformly random sample form the solution space in time $O(n \log(1/\eta))$ with probability $1 - \eta$.*

Note that when the algorithm fails, it does not output a solution. Any solutions it outputs are guaranteed to be distributed uniformly.

*Proof.* We set $\delta = 0.1$ and construct $M^0$ which requires time $O(n^3 \log(W) \log(n))$. It is easy to see that $M^i(v, z) \leq M(v, z)$ for any vertex $v \in M^i$ and $i \leq n$, hence $A_M(v) \subseteq A_{M^i}(v)$ and in particular $A_M(s) \subseteq A_{M^0}(s)$. By Lemma 3.3, $|A_{M^0}(s)| \leq 1.1 |A_M(s)|$.

Further it is easy to sample from $A_{M^0}(s)$ in time $O(n)$. Recall we have $P_{M^0}(v)$ computed for each state $v$. We start at $s$. From a current vertex $v$, we move to $v_b = M^0(v, b)$ for $b \in \{0, 1\}$ with probability $\frac{P_{M^0}(v_b)}{P_{M^0}(v_0) + P_{M^0}(v_1)}$. This produces $z \in_u A_{M^0}(s)$. We check that $z$ is also a solution to the original knapsack in time $O(n)$, this happens with probability at least $0.8$. By repeating this $O(\log(\eta^{-1}))$ times, the failure probability becomes less than $\eta$. $\square$

## 4 Monotone ROBPs, Small-Space Sources, and Counting Solutions to Multidimensional Knapsack

In this section, we consider more general models of computation and wider classes of distributions. We solve the approximate counting problem for the more general class of *monotone* read-once branching programs as defined in the work of Meka and Zuckerman [MZ10]. Further, we show how to deterministically approximate the acceptance probability under the natural and broader class of *small-space sources* introduced by Kamp et al. [KRVZ06].

Monotone ROBPs were introduced by Meka and Zuckerman [MZ10] in their work on pseudo-random generators for halfspaces. In addition to halfspaces, the class of monotone ROBPs includes read-once DNFs and read-once polynomial threshold functions (read-once PTFs).

**Definition 4.1** (Monotone ROBP). *A $(W, n)$-branching program $M$ is said to be monotone if for all $i \leq n$, there exists a total ordering $\prec$ on the vertices in $L(M, i)$ such that if $u \prec v$, then $A_M(u) \subseteq A_M(v)$.*

It is easy to see that the branching program for knapsack satisfies the above condition. Given partial sums $v_j, v_k$ we say $v_j \prec v_k$ if $v_j > v_k$, since a larger partial sum means that fewer suffix strings will be accepted. We say $u \preceq v$ if $u \prec v$ or $u = v$.

Since we deal with monotone ROBPs that potentially have width exponential in $n$, we require that $M$ is described implicitly in the following sense:

1. Ordering: given two states $u, v$ we can efficiently check if $u \prec v$ and if so find a $w$ that is half-way between $u, v$, i.e., $||\{x : u \prec x \prec w\}| - |\{x : w \prec x \prec v\}|| \leq 1$.

2. Transitions: Given any vertex of $M$ we can compute the two neighbors of the vertex.

We assume that the above two operations are of unit cost.

Our counting result for monotone ROBPs is obtained by proving the following structural result for monotone ROBPs that we believe is of independent interest:

**Theorem 4.2** (Main). *Given a $(W, n)$-monotone ROBP $M$, $\delta > 0$, and a small-space distribution $D$ over $\{0, 1\}^n$ of width at most $S$, there exists an $(O(n^2 S/\delta), n)$-monotone ROBP $M^0$ such that for all $z$, $M(z) \leq M^0(z)$ and*

$$\Pr_{x \leftarrow D}[M(z) = 1] \leq \Pr_{x \leftarrow D}[M^0(z) = 1] \leq (1 + \delta) \Pr_{x \leftarrow D}[M(z) = 1].$$

*Moreover, given an implicit description of $M$ and an explicit description of $D$, $M^0$ can be constructed in deterministic time $O(n^3 S(S + \log(W)) \log(n/\delta)/\delta)$.*

We prove Theorem 4.2 in Section 4.2. As discussed in the introduction, this theorem has many interesting consequences. We first derive these consequences before proving the theorem. Theorem 1.6 follows easily from the observation that a weight $W$ halfspace is a $(W, n)$-monotone ROBP. Corollary 1.7 follows since the uniform distribution over strings of weight exactly $r$ can be generated by a small space source of width at most $r + 1$. Further we can approximately count knapsack solutions with respect to all symmetric distributions, and all product distributions, since each of these can be generated by a small space source.

### 4.1 A FPTAS for Multidimensional Knapsack

Combining Theorem 4.2 work with a result due to Dyer [Dye03], we obtain a deterministic approximate counting algorithm for multi-dimensional knapsack with a constant number of constraints, matching Dyer's FPRAS up to polynomial factors. We use the following elegant rounding result due to Dyer:

**Theorem 4.3** (Dyer, [Dye03]). *Given knapsack instances $\mathsf{KNAP}(a_1, b_1), \ldots, \mathsf{KNAP}(a_k, b_k)$, we can deterministically in time $O(n^3(\log W))$ construct a new set of knapsack instances $\mathsf{KNAP}(a'_1, b'_1), \ldots, \mathsf{KNAP}(a'_k, b'_k)$ each with a total weight of at most $O(n^3)$ such that $\mathsf{KNAP}(a_i, b_i) \subseteq \mathsf{KNAP}(a'_i, b'_i)$, $\forall 1 \leq i \leq k$, and*

$$|\cap_i \mathsf{KNAP}(a'_i, b'_i)| \leq (n + 1)^k |\cap_i \mathsf{KNAP}(a_i, b_i)|.$$

*Proof of Theorem 1.2.* We first use Dyer's algorithm to obtain low-weight knapsack instances $\mathsf{KNAP}(a_1', b_1'), \cdots, \mathsf{KNAP}(a_k', b_k')$ as in Theorem 4.3. Let $D$ be the uniform distribution over the set $U = \cap_i \mathsf{KNAP}(a_i', b_i')$ and observe that by Corollary 2.4 $D$ can be generated by an explicit $O(n^{3k})$ space source. For $i \in [k]$, let $M^i$ be a $(W, n)$-ROBP exactly computing the indicator function for $\mathsf{KNAP}(a_i, b_i)$. Let $\delta = O(\varepsilon/k(n+1)^k)$ to be chosen later. Now, for every $i \in [k]$, by Theorem 4.2 we can explicitly in time $n^{O(k)}(\log W)/\delta$ construct a $(n^{O(k)}/\delta, n)$-ROBP $M^i_{up}$ such that,

$$\Pr_{x \leftarrow D}[M^i_{up}(x) \neq M^i(x)] \leq \delta.$$

Let $M$ be the $(n^{O(k^2)}/\delta^k, n)$-ROBP computing the intersection of $M^i_{up}$ for $i \in [k]$, i.e., $M(x) = \wedge_i M^i_{up}(x)$. Then, by a union bound,

$$\Pr_{x \leftarrow D}[M(x) \neq \wedge_i M^i(x)] \leq k\delta.$$

On the other hand, by Theorem 4.3,

$$\Pr_{x \leftarrow D}[\wedge_i M^i(x) = 1] \geq 1/(n+1)^k.$$

Therefore, from the above two equations and setting $\delta = \varepsilon/2k(n+1)^k$, we get that

$$\Pr_{x \leftarrow D}[M(x) = 1] \leq \Pr_{x \leftarrow D}[\wedge_i M^i(x) = 1] \leq (1+\varepsilon) \Pr_{x \leftarrow D}[M(x) = 1].$$

Thus, $p = \Pr_{x \in_u \{0,1\}^n}[x \in U] \cdot \Pr_{x \leftarrow D}[M(x) = 1]$ is an $\varepsilon$-relative error approximation to the fraction of solutions to all constraints $\Pr_{x \in_u \{0,1\}^n}[\wedge_i M^i(x) = 1] = \Pr_{x \in_u \{0,1\}^n}[x \in U] \cdot \Pr_{x \leftarrow D}[\wedge_i M^i(x) = 1]$.

The theorem now follows since we can compute $p$ in time $(n/\delta)^{O(k^2)}$ using Claim 2.3, as $D$ is a small-space source of width at most $O(n^{3k})$ and $M$ has width at most $(n/\delta)^{O(k^2)}$. $\square$

## 4.2 Proof of Theorem 4.2

We start with some notation. Let $D$ denote the small space generator of width at most $S$. For $A \subseteq \{0,1\}^n$ we use $D(A)$ to denote the measure of $A$ under $D$. Let $U^1, \ldots, U^n$ be the vertices in $D$ with $U^i$ being the $i$'th layer of $D$. For a vertex $u \in U^i$, let $D^u$ be the distribution over $\{0,1\}^{n-i}$ induced by taking a random walk in $D$ starting from $u$. Given a vertex $v \in L(M, i)$ and $u \in U^i$, let $P_{M,u}(v)$ denote the probability of accepting if we start from $v$ and make transitions in $M$ according to a suffix sampled from distribution $D^u$.

As we did for knapsack, we start from the exact branching program $M$ and construct a sequence of programs $M^n = M, \ldots, M^0$, where $M^i$ is obtained from $M^{i+1}$ by rounding the $(i+1)^{st}$ layer. We do the rounding in such a way that the acceptance probabilities are well approximated under each of the possible distributions on suffixes $D^u$. The program $M^0$ will be a small width program.

Let $L(M^{i+1}, i+1) = \{v_1 \prec v_2 \cdots \prec v_W\}$. Fix a vertex $u \in U^{i+1}$. We define a set $B^{i+1}(u) = \{v_{u(j)}\} \subseteq L(M^{i+1}, i+1)$ of breakpoints for $u$ as follows. We start with $v_{u(1)} = v_W$ and given $v_{u(j)}$ define $v_{u(j+1)}$ by

$$v_{u(j+1)} = \max v \ s.t. v \prec v_{u(j)} \text{ and } 0 < P_{M^{i+1}, u}(v) < P_{M^{i+1}, u}(v_{u(j)})/(1+\varepsilon) \tag{4.1}$$

Let $B^{i+1} = \cup_{u \in U^{i+1}} B^{i+1}(u) = \{b_1 \prec \cdots \prec b_N\}$ be the union of breakpoints for all $u$. We set $L(M^i, i+1) = B^{i+1}$. The vertices in all other layers stay the same as in $M^{i+1}$, as do all the edges except those from layer $i$ to $i+1$. We round these edges *upward* as follows: let $v \in L(M^{i+1}, i)$ and $M^{i+1}(v, b) = v' \in L(M^{i+1}, i+1)$. Find two consecutive vertices $b_k, b_{k+1} \in L(M^i, i+1)$ such that $b_k \prec v' \preceq b_{k+1}$. We set $M^i(v,b) = b_{k+1}$. Note that this only increases the number of accepting suffixes for $v$.

This completes the construction of the $M^i$s. We now analyze the running time of our algorithm. We start with the following claims whose proofs are similar to that of Lemmas 3.1, 3.3 and are omitted.

**Lemma 4.4.** *The branching program $M^i$ is monotone where the ordering of vertices in each layer is the same as $M$.*

**Lemma 4.5.** *For $v \in M^i$, we have $A_{M^{i+1}}(v) \subseteq A_{M^i}(v)$. Thus $P_{M,u}(v) \leq P_{M^i,u}(v)$ for all $u \in U^i$.*

We next analyze the complexity of constructing $M^0$ from $M$.

**Lemma 4.6.** *The branching program $M^0$ can be constructed in time $O(n^2 S(S+\log(W))\log(nS/\varepsilon)/\varepsilon)$.*

*Proof.* Observe that for every $i$ and $u \in U^i$, $|B^i(u)| \leq \frac{2n}{\varepsilon}$ and hence $|B^i| \leq \frac{2nS}{\varepsilon}$. Let us analyze the complexity of constructing $M^i$ from $M^{i+1}$. We will assume inductively that the set $B^{i+2}$ is known and stored in a binary tree along with the values $P_{M^{i+1},u}(b)$, for every $b \in B^{i+2}$ and $u \in U^{i+2}$. Hence, given $v \in L(M, i+1)$, we can find $b_k, b_{k+1} \in B^{i+2}$ such that $b_k \prec v \preceq b_{k+1}$ in time $\log(nS/\varepsilon)$. This ensures that if we are given a vertex $v' \in L(M^{i+1}, i+1)$ and $u \in U^{i+1}$, we can compute $P_{M^{i+1},u}(v')$ in time $\log(nS/\varepsilon)$. To see this, note that

$$P_{M^{i+1},u}(v') = \sum_{z \in \{0,1\}} p_u(z) P_{M^{i+1},u_z}(M^{i+1}(v', z))$$

where $u_z \in U^{i+2}$ denotes the vertex reached in $D$ when taking the edge labeled $z$ from $u$. To compute $M^{i+1}(v', z)$ we first compute $v = M(v', z)$ using the fact that $M$ is described implicitly. We then find $b_k \prec v \preceq b_{k+1}$ in $B^{i+2}$ and set $M^{i+1}(v', z) = b_{k+1}$. Since we have the values of $P_{M^{i+1},u_z}(b)$ precomputed, we can use them to compute $P_{M^{i+1},u}(v')$. The time required is dominated by the $O(\log(nS/\varepsilon))$ time needed to find $b_{k+1}$.

Now, for each $u \in U^{i+1}$, by using binary search on the set of vertices as in Lemma 3.2, each new breakpoint in $B^{i+1}(u)$ can be found in time $O(\log(W)\log(nS/\varepsilon))$. Thus finding the set $B^{i+1}$ takes time $O(nS \log(W)\log(nS/\varepsilon)/\varepsilon)$.

Once we find the set $B^{i+1}$, we store it as a binary tree. We compute and store the values of $P_{M^i,u}(b) = P_{M^{i+1},u}(b)$ for each $b \in B^{i+1}$ and $u \in U^{i+1}$ in time $O(nS^2 \log(nS/\varepsilon)/\varepsilon)$.

Thus overall, the time required to construct $M^0$ from $M$ is $O(n^2 S(S+\log(W))\log(nS/\varepsilon)/\varepsilon)$. □

We next show that the number of accepting solutions does not increase by too much.

**Lemma 4.7.** *For $v \in L(M^i, j)$ and $u \in U^j$, we have $P_{M^i,u}(v) \leq P_{M,u}(v)(1+\varepsilon)^{n-i}$.*

*Proof.* It suffices to show that $P_{M^i,u}(v) \leq P_{M^{i+1},u}(v)(1+\varepsilon)$. This claim is trivial for $j \geq i+1$ since for such vertices, $P_{M^i,u}(v) = P_{M^{i+1},u}(v)$. As in Lemma 3.4, the crux of the argument is when $j = i$. Since $M^{i+1}$ and $M^i$ are identical up to layer $i$, the claim for $j < i$ will follow.

Fix $v \in L(M^i, i)$ and $u \in U^i$. Let $u_0, u_1$ denote the neighbors of $u$ in $D$. Then we have

$$P_{M^i,u}(v) = p_u(0) P_{M^i,u_0}(M^i(v,0)) + p_u(1) P_{M^i,u_1}(M^i(v,1)). \tag{4.2}$$

We first bound $P_{M^i,u_0}(M^i(v,0))$. Let $b_1, b_4$ be the breakpoints in $B^{i+1}(u_0)$ such that $b_1 \prec M^{i+1}(v,0) \preceq b_4$ and let $b_2, b_3$ be the breakpoints in $B^{i+1}$ such that $b_2 \prec M^{i+1}(v,0) \preceq b_3$. Note that $M^i(v,0) = b_3$, by the construction of $M^i$. Since $B^{i+1}(u_0) \subseteq B^{i+1}$, we get $b_1 \preceq b_2 \prec M^{i+1}(v,0) \preceq b_3 \preceq b_4$. By the definition of breakpoints, we have

$$P_{M^{i+1},u_0}(b_4) \leq (1+\varepsilon)P_{M^{i+1},u_0}(M^{i+1}(v,0))$$

and by the monotonicity of $M^{i+1}$

$$P_{M^{i+1},u_0}(b_3) \leq P_{M^{i+1},u_0}(b_4).$$

which together show that

$$P_{M^{i+1},u_0}(b_3) \leq (1+\varepsilon)P_{M^{i+1},u_0}(M^{i+1}(v,0)).$$

Since $b_3 \in L(M^i, i+1)$, we have $P_{M^i,u_0}(b_3) = P_{M^{i+1},u_0}(b_3)$. Thus

$$P_{M^i,u_0}(M^i(v,0)) \leq (1+\varepsilon)P_{M^{i+1},u_0}(M^{i+1}(v,0)).$$

Similarly, we can show

$$P_{M^i,u_1}(M^i(v,1)) \leq (1+\varepsilon)P_{M^{i+1},u_1}(M^{i+1}(v,1)).$$

Plugging these into Equation 4.2 gives

$$P_{M^i,u}(v) \leq (1+\varepsilon)(p_u(0)P_{M^{i+1},u_0}(M^{i+1}(v,0)) + p_u(1)P_{M^{i+1},u_1}(M^{i+1}(v,1))) = (1+\varepsilon)P_{M^{i+1},u}(v)$$

which is what we set out to prove. □

We can now prove Theorem 4.2.

*Proof.* Choose $\varepsilon = \Omega(\delta/n)$ so that $(1+\varepsilon)^n \leq (1+\delta)$. We construct the program $M^0$ from $M$ and output $P_{M^0,u}(s)$ where $s$ is the start state of $M$ and $u$ is the start state of $S$. By Lemma 4.6, this takes time $O(n^3 S(S + \log(W)) \log(nS/\delta)/\delta)$. Applying Lemmas 4.7 and 4.5, we conclude that

$$P_{M,u}(s) \leq P_{M^0,u}(s) \leq P_{M,u}(s)(1+\delta).$$

Note that $P_{M,u}(s)$ and $P_{M^0,u}(s)$ are respectively the probabilities that $M$ and $M^0$ accept a string sampled from the distribution $D$. This completes the proof. □

## 5 Counting for General Integer Knapsack and Contingency Tables

Our algorithms for counting also extend to general integer knapsack and contingency tables. Conceptually the algorithms are similar to those for $\{0,1\}$-knapsack and multidimensional knapsack. However, the details are a little intricate involving a combination of our ideas and Dyer's ideas. We defer the proofs to the appendix.

## 6 Learning Functions of Halfspaces via ROBPs

We now present our learning algorithm and prove Theorem 1.5. We start with some notation. A halfspace $h : \{0,1\}^n \to \{0,1\}$ is a Boolean function defined by $f(x) = 1$ if $\sum_i a_i x_i \leq b$ and 0 otherwise, where $a \in \mathbb{R}^n$ and $b \in \mathbb{R}$. Let $f : \{0,1\}^n \to \{0,1\}$ and let $\mu_i$ denote the uniform distribution over $\{0,1\}^i$. For each prefix $x \in \{0,1\}^i$, we define the function $f_x : \{0,1\}^{n-i} \to \{0,1\}$ by $f_x(z) = f(x \circ z)$ where $\circ$ denotes concatenation. For two functions $f, g : \{0,1\}^n \to \{0,1\}$, we define $d(f,g) = \Pr_{x \leftarrow \mu_n}[f(x) \neq g(x)]$. Thus given two prefixes $x, y \in \{0,1\}^i$, $d(f_x, f_y) = \Pr_{z \leftarrow \mu_{n-i}}[f(x \circ z) \neq f(y \circ z)]$.

**Definition 6.1** (Almost ROBPs). *We call a function $f : \{0,1\}^n \to \{0,1\}$ a $(\varepsilon, W, n)$-almost ROBP if there exist sets $S^l \subseteq \{0,1\}^l$, $l \in \{1, \ldots, n\}$ with $|S^l| \leq W$ such that for every $y \in \{0,1\}^l$ there exists an $x \in S^l$ such that $d(f_x, f_y) \leq \varepsilon$. We call sets $S^1, \ldots, S^l$ $(\varepsilon, W)$-representatives.*

It is interesting to contrast the notion of an almost-ROBP (aROBP for short) with having a good approximation by an ROBP under the uniform distribution. It is easy to show that there exists a function $f$ which is $\delta$-close to a width 2 ROBP, but which is not an $(\varepsilon, W, n)$-aROBP for $W, \varepsilon^{-1} = \text{poly}(n, \delta^{-1})$, by corrupting the parity function randomly on some $\delta$ fraction of inputs. In the other direction, it is not obvious that an $(\varepsilon, W, n)$-aROBP can be well-approximated by a small width ROBP under the uniform distribution. But this is in fact true, and the proof is via our learning algorithm, which we present below.

The algorithm learns an aROBP $f$, given query access to $f$, by constructing a ROBP $M$ that approximates $f$. The ROBP $M$ has $n$ layers numbered 0 through $n$. The set of vertices in layer $i$ is denoted by $L(M, i)$. Each vertex $x \in L(M, i)$ corresponds to a string $x \in \{0,1\}^i$. $L(M, 0)$ consist of a single start state, identified with the null string $\varphi$. By abuse of notation, we will think of $M$ both as a branching program and a Boolean function.

---

**Main Algorithm**. Input $n, \varepsilon, W$.

Let $L(M, 0)$ contain the null string, while $L(M, i)$ are empty sets for $i \in \{1, \ldots, n\}$.
For $i = 1, \ldots, n$:
  For each $x \in L(M, i-1)$ and $b \in \{0, 1\}$,
    Check if there is $y \in L(M, i)$ such that $d(f_{x \circ b}, f_y) \leq 3\varepsilon$.
      If so, add an edge labeled $b$ from $x$ to $y$.
      If not, add $x \circ b$ to $L(M, i)$, add an edge labeled $b$ from $x$ to $x \circ b$.
  If $|L(M, i)| > W$, then output FAIL and halt.

---

In line 4 of our algorithm, to check if there is a vertex $y$ that is $\varepsilon$-close to $x \circ b$, we pick $L$ random suffixes $z \in \{0,1\}^{n-i}$ and check if $f(x \circ b \circ z) = f(y \circ z)$. By the Chernoff bound, if $L = O(\log(nW^2/\delta)/\varepsilon)$, then the probability that our estimate of $d(f_{x \circ b}, f_y)$ is off by more than an additive $\varepsilon$ is at most $\delta/2nW^2$. Since each layer has at most $W$ vertices in total, we estimate at most $2nW^2$ such quantities. Hence the probability that the error is more than $\varepsilon$ in any of our estimates is at most $\delta$.

**Theorem 6.2.** *For $\varepsilon, \delta > 0$, given oracle access to a $(\varepsilon, W, n)$-almost ROBP $f$, the above algorithm runs in time $O(nW \log(nW/\delta)/\varepsilon)$ and constructs a $(W, n)$-ROBP $M$ such that $d(M, f) \leq 4n\varepsilon$ with probability at least $1 - \delta$.*

We assume that all our estimates are within $\varepsilon$, which happens with probability $1 - \delta$. The theorem follows from two claims.

**Claim 6.3.** *The algorithm never outputs FAIL.*

*Proof.* Let $S^1, \ldots, S^n$ be $(\varepsilon, W)$-representatives for $f$. For each $x \in S^i$, consider the balls $B(x) = \{y \in \{0,1\}^i : d(f_y, f_x) \leq \varepsilon\}$ for any $x \in S^i$. By definition, they cover all of $\{0,1\}^{n-i}$. We claim that $L(M, i)$ cannot have two distinct vertices $y, y' \in \{0,1\}^i$ in layer $i$ that belong to the same ball $B(x)$. For, if $y, y'$ lie in the same ball, $d(f_y, f_{y'}) \leq 2\varepsilon$. Since the sampling error is at most $\varepsilon$, our estimate for $d(f_y, f_{y'})$ would be at most $3\varepsilon$, thus we would not add both of them to $L(M, i)$. Hence $|L(M, i)| \leq |S^i| \leq W$. □

**Claim 6.4.** *We have $\Pr_\mu[M(x) \neq f(x)] \leq 4n\varepsilon$.*

*Proof.* By induction on $n - i$, we will show that for every $x \in L(M, i)$, $d(M_x, f_x) \leq 4(n-i)\varepsilon$. This implies that when $i = 0$, $d(M, f) \leq 4n\varepsilon$ as desired.

For $i = n$ there is nothing to prove. Suppose the statement is true for all vertices in $L(M, i+1)$. Consider a vertex $x \in L(M, i)$. Let $y_0, y_1 \in L(M, i+1)$ be it's neighbors in $M$. Then, by our assumption on sampling errors, for $b \in \{0, 1\}$, $d(f_{x \circ b}, f_{y_b}) \leq 4\varepsilon$. By the induction hypothesis, we know that $d(f_{y_b}, M_{y_b}) \leq 4(n - i - 1)\varepsilon$. Putting these together, we get

$$\begin{aligned}
d(f_x, M_x) &= \frac{1}{2} \sum_{b \in \{0,1\}} d(f_{x \circ b}, M_{y_b}) \\
&\leq \frac{1}{2} \sum_{b \in \{0,1\}} (d(f_{x \circ b}, f_{y_b}) + d(f_{y_b}, M_{y_b})) \quad \text{(by triangle inequality)} \\
&\leq 4\varepsilon + 4(n - i - 1)\varepsilon = 4\varepsilon(n - i).
\end{aligned}$$

□

Theorem 6.2 now follows as the probability of sampling error is at most $\delta$. Our main learning result for halfspaces, Theorem 1.5 follows by combining Theorem 6.2 and the following easy claims. The first claim is implicit in [MZ10] who prove a stronger result about sandwiching halfspaces between ROBPs. We present a more direct proof below.

**Claim 6.5.** *Every halfspace is an $(\varepsilon, 1/\varepsilon, n)$-almost ROBP.*

*Proof.* Fix a halfspace $f \equiv 1\{\sum_i a_i x_i \leq b\}$. Fix $i \leq n$. We show that there exist representatives $S^i$, $|S^i| \leq 1/\varepsilon$ for prefixes of length $i$. Let $g(v) = \Pr_\mu[\sum_{j \leq i} a_i x_i \leq v]$. Observe that $g(v)$ is a non-decreasing function of $v$. Now, starting from $v_1 = 0$ we inductively define $v_{j+1} = \min v > v_j$ such that $g(v) \geq g(v_j) + \varepsilon$. This gives at most $k \leq 1/\varepsilon$ values $v_j$. Now for each $j$, we choose $x_j$ to be some $x \in \{0,1\}^i$ such that $\sum_{l \leq i} a_l x_l = v_j$. It is easy to see that $S^i = \{x_1, \ldots, x_k\}$ forms a set of representatives for prefixes of length $i$. □

**Claim 6.6.** *Let $f^1, \ldots, f^k : \{0,1\}^n \to \{0,1\}$ be $(\varepsilon, W, n)$-aROBPs and $g : \{0,1\}^k \to \{0,1\}$. Then $h : \{0,1\}^n \to \{0,1\}$ defined by $h(x) = g(f^1(x), \ldots, f^k(x))$ is an $(2k\varepsilon, W^k, n)$-aROBP.*

*Proof.* For $j \leq k$, let $S_j^1, \ldots, S_j^n$ be $(\varepsilon, W)$-representatives for $f_j$. Fix $i \leq n$ and form a set of prefixes $T^i \subseteq \{0,1\}^i$ as follows: for every $x_1 \in S_1^i, \ldots, x_k \in S_k^i$ if $U(x_1, \ldots, x_k) = \{z \in \{0,1\}^i : d(f_{x_j}^j, z) \leq \varepsilon, \forall j \leq k\}$ is not empty, add a single element of $U$ to $T^i$.

By construction, $|T^i| \leq W^k$. Further, for every $y \in \{0,1\}^i$, there exist $x_1 \in S_1^i, \ldots, x_k \in S_k^i$ such that $y \in U(x_1, \ldots, x_k)$. Let $u$ be the element of $U(x_1, \ldots, x_k)$ added to $T$. Then, by a union bound, $d(h_y, h_u) \leq \sum_j d(f_y^j, f_u^j) \leq 2k\varepsilon$. □

We observe that combing the above arguments with those of Theorem 4.2, we get similar results for learning under any explicitly given small-space source. In particular we can learn functions of halfspaces under $p$-biased and symmetric distributions.

# A  A FPTAS for General Integer Knapsack

In this section, we prove Theorem 1.3. As in the case of $\{0,1\}$-knapsack we start with the exact branching program $M$ for $\mathsf{KNAP}(a,b,u)$, where each state in $L(M,j)$ corresponds to a partial sum $v_j = \sum_{i \leq j} a_i x_i$ and has $(u_{j+1}+1)$ outgoing edges corresponding to the possible values of variable $x_{j+1}$. We then approximate this program with a small width branching program. However the program $M$ can both large width and large degree. To handle this, we observe that the branching program $M$ is an *interval ROBP* in the sense defined below, which allows us to shrink the state space, and obtain succinct descriptions of the edges of the new branching program we construct.

**Definition A.1** (Interval ROBPs). *For $u = (u_1, \ldots, u_n) \in \mathbb{Z}_+^n$, $S, T \in \mathbb{Z}_+$, an $(S, u, T)$-interval ROBP $M$ is a layered multi-graph with a layer for each $0 \leq i \leq T$, at most $S$ states in each layer. The first layer has a single (start) vertex, each vertex in the last layer is labeled accepting or rejecting and there exists a total order $\prec$ on the vertices of layer $i$ for $0 \leq i \leq T$. A vertex $v$ in layer $i-1$ has exactly $u_i + 1$ edges labeled $\{0, 1, \ldots, u_i\}$ that respect the ordering $\prec$: If $M(v, k)$ denotes the k'th neighbor of $v$ for $k \leq u_i$, then $M(v, u_i) \preceq M(v, u_i - 1) \preceq \cdots \preceq M(v, 0)$.*

*An interval ROBP defines a function $M : [0, u_1] \times [0, u_2] \times \cdots \times [0, u_n] \to \{0, 1\}$ where on input $x$, we begin at the start vertex and output the label of the final vertex reached when traversing $M$ according to $x$.*

Given an $(S, u, T)$-interval ROBP $M$, and a vertex $v \in L(M, i-1)$, the edges out of $v$ can be described succinctly by a subset of at most $S$ edges irrespective of how large $u_i$ is. If we set $E(v, w) = \{0 \leq k \leq u_i : M(v, k) = w\}$ be the set of edges from $v$ to $w$, then $E(v, w)$ is an interval, meaning $E(v, w) = \{l_{v,w}, \ldots, r_{v,w}\}$ for some $l_{v,w}, r_{v,w} \in \mathbb{Z}_+$. Thus, to describe $E(v, w)$ we only need to know $l_{v,w}$ and $r_{v,w}$. This allows us to succinctly describe the interval ROBP $M^0$ approximating $\mathsf{KNAP}(a, b, u)$ by storing just the end points of the edge sets $E(v, w)$ for $v, w \in M^0$.

Let $M$ denote the exact branching program for $\sum_{i \leq n} a_i x_i \leq b$ with edges between layers $i-1$ and $i$ labeled by $x_i \in \{0, \ldots, u_i\}$ and for $v \in L(M, i-1)$, $0 \leq x_i \leq u_i$ we have $M(v, x_i) = v + a_i x_i \in L(M, i)$. Given a vertex $v \in L(M, j)$ we use $P_M(v)$ to denote the probability that $M(v, z)$ accepts, for $z$ chosen randomly form $\{0, \ldots, u_{j+1}\} \times \cdots \times \{0, \ldots, u_n\}$. It is clear from the definition that $M$ is an interval ROBP with the ordering on $L(M, i)$ given by $u \prec v$ if $u > v$.

We construct a series of interval ROBPs $M^n = M, M^{n-1}, \ldots, M^0$. We obtain $M^i$ from $M^{i+1}$ by rounding the states in $L(M^{i+1}, i+1)$. More precisely, we set $L(M^i, i+1) = \{v_1, \ldots, v_\ell\} \subseteq L(M^{i+1}, i+1)$ where the $v_j$s are defined as follows: Let $v_1 = 0$. Given $v_j$, let

$$v_{j+1} = \min v \text{ such that } v > v_j \text{ and } 0 < P_{M^{i+1}}(v) < P_{M^{i+1}}(v_j)/(1+\eta). \tag{A.1}$$

Let $I_1 = \{v_1, \ldots, v_2 - 1\}, \ldots, I_\ell = \{v_\ell, \ldots\}$, where $\ell \leq n(\log U)/\eta$ as $P_{M^i}(v_1) \leq 1$ and $P_{M^i}(v_\ell) \geq U^{-n}$. Next we redirect the transitions going from level $i$ to level $i+1$. If we have an edge labeled $z \in \{0, \ldots, u_{i+1}\}$ entering a vertex $v \in I_j$, then we redirect the edge to vertex $v_j$. The redirection will be done implicitly in the sense that for any vertex $v$ in level $i$ and a vertex $v_j$, we only compute and store the end points of the interval $E(v, v_j) = \{0 \leq k \leq u_{i+1} : M^i(v, k) = v_j\}$.

Our branching programs have the following approximating properties analogous to Lemmas 3.1, 3.3, 3.4. The proofs are similar and are omitted.

**Lemma A.2.** *For any $v \in L(M^i, j)$ and $0 \leq k \leq l \leq u_{j+1}$, $M^i(v, k) \leq M^i(v, l)$. Let $v, v' \in L(M^i, j)$ and $v \leq v'$. For any suffix $z$, $M^i(v, z) \leq M^i(v', z)$.*

**Lemma A.3.** *For $v \in M^i$, we have $A_{M^{i+1}}(v) \subseteq A_{M^i}(v)$. Further, for any $v \in L(M^i, j)$ where $j \leq i$, we have $P_M(v) \leq P_{M^i}(v) \leq P_M(v)(1+\eta)^{n-i}$.*

We next show that $M^0$ can be constructed efficiently.

**Lemma A.4.** *Each vertex $v_j \in L(M^i, i+1)$ can be computed in time $O(n(\log U)(\log W)/\eta)$.*

*Proof.* The proof is by induction: we maintain the invariant that for every $i$, we know the vertices $v_j$ of $L(M^i, i+1)$ and their acceptance probabilities $P_{M^i}()$.

Suppose we have the above setup for $l > i$ and have computed $v_1, \ldots, v_j \in L(M^i, i+1)$. Recall that $v_{j+1}$ is the smallest value of $v > v_j$ satisfying $P_{M^{i+1}}(v) < P_{M^{i+1}}(v_j)/(1+\eta)$. We next show that for a given $v \in L(M^{i+1}, i+1)$, $P_{M^{i+1}}(v)$ can be computed in time $O(n(\log U)/\eta)$. Let $L(M^{i+1}, i+2) = \{w_1 < w_2 < \cdots\}$ then, $E(v, w_l) = \{0 \le k \le u_{i+2} : w_l \le v + a_{i+2}k < w_{l+1}\}$ and

$$P_{M^{i+1}}(v) = \sum_{w \in L(M^{i+1}, i+2)} \frac{|E(v, w)|}{u_{i+2} + 1} P_{M^{i+1}}(w).$$

Thus, we can compute $P_{M^{i+1}}(v)$ in time $O(n(\log U)/\eta)$ as $|L(M^{i+1}, i+2)| \le n(\log u)/\eta$.

We can now do binary search on $P_{M^{i+1}}(v)$. Since we start with integers in the range $\{0, \ldots, W\}$, this will require $O(\log(W))$ computations of $P_{M^{i+1}}(v)$. Once we have computed $L(M^i, i+1)$ we store these vertices and their probabilities of acceptance. □

Thus we can construct $M^0$ from $M$ in time $O(n^3(\log U)^2(\log W)/\eta^2)$.

We can now finish the proof of our counting result for general integer knapsack.

*Proof of Theorem 1.3.* We set $\eta = \delta/2n$ and use the above arguments to construct the branching program $M^0$ and compute the value of $P_{M^0}(s)$ where $s$ is the start state. We now apply Lemma A.3 to conclude that

$$P_M(s) \le P_{M^0}(s) \le P_M(s)(1+\eta)^n \le (1+\delta)P_M(s)$$

where the last inequality holds for small enough $\delta$. Finally, note that the number of knapsack solutions is precisely $P_M(s) \prod_i (u_i + 1)$. Hence we output $P_{M^0}(s) \prod_i (u_i + 1)$. □

## B  A Deterministic Algorithm for Counting Contingency Tables

We now address the question of counting contingency tables. Our algorithm is fairly intricate and involves a combination of Dyer's FPRAS for counting contingency tables and our algorithms for counting general integer knapsack solutions and counting knapsack solutions under small space sources. Here is a high-level outline of the algorithm:

- We first give an algorithm for counting integer knapsack solutions under "interval small-space sources" which are integer-valued distributions that generalize small-space sources in the same vein as interval ROBPs of Definition A.1 generalize ROBPs. However, we specialize our analysis to the specific case of contingency tables for clarity.

- We then observe that Dyer's approach for counting contingency tables (implicitly) gives an explicit "interval small-space source" $D$ whose support contains all feasible contingency tables and the set of feasible contingency tables has non-negligible density under $D$. We then combine the above two observations as in the proof of Theorem 1.2.

We first set up some notation. Following Dyer, we will solve the following formulation of counting contingency tables. Given $r = (r_1, \ldots, r_m) \in \mathbb{Z}_+^m$, $c = (c_1, \ldots, c_n) \in \mathbb{Z}_+^n$, estimate $|CT'(r,c)|$, where

$$CT'(r,c) = \{X \in \mathbb{Z}_+^{m \times n} : \sum_{j=1}^n X_{ij} \leq r_i, i \in [m], \sum_{i=1}^m X_{ij} = c_j, j \in [n]\}.$$

This is equivalent to the original problem as stated in the introduction, since for $r \in \mathbb{Z}_+^m$ and $c' = (c_1, \ldots, c_{n+1}) \in \mathbb{Z}_+^{n+1}$, $|CT(r,c')| = |CT'(r,c)|$, where $c = (c_1, \ldots, c_n)$. For $X \in \mathbb{Z}_+^{m \times n}$, let $X^i$ denote the $i$'th row and $X_j$ denote the $j$'th column of $X$. Let $R = \max(r_i, c_j : i \in [m], j \in [n])$. For $j \in [n]$, let $B(j) = \{x \in \mathbb{Z}_+^m : \sum_i x_i = c_j\}$ and let $B = \{X : X_j \in T(j), j \in [n]\}$. Further, define $h : \mathbb{Z}_+^m \to \{0, \ldots, 2n^2\}^m = T$ by $h_i(x) \equiv (h(x))_i \equiv \lfloor 2n^2 x_i / r_i \rfloor$ and let

$$S = \{X \in B : \sum_{j=1}^n h(X_j) \leq 2n^2 \mathbf{1}\}.$$

We now state a sequence of lemmas that we need for our proof. The first is due to Dyer [Dye03]. At a high level, it lets us use the uniform distribution over $S$ in the role of a small-space source.

**Lemma B.1** (Dyer). *$CT'(r,c) \subseteq S$ and $|S| \leq n^m |CT'(r,c)|$. Further, we can estimate $|S|$ deterministically in time $n^{O(m)}$.*

Given this lemma, it suffices to additively approximate the number of valid contingency tables under $S$, which we do by constructing suitable ROBPs. The next lemma gives us explicit small-width interval ROBPs $M_i$ that approximate the $i$'th row constraint under the uniform distribution over $S$.

**Lemma B.2.** *For every $i \in [m]$, we can in time $n^{O(m)}(\log^3 R)/\eta^2$ compute an $(n^{O(m)}(\log R)/\eta, c, n)$-interval ROBP $M_i$ explicitly such that for every $x \in \mathbb{Z}_+^n$, $\sum_j x_j \leq r_i$ implies $M_i(x) = 1$, and*

$$\Pr_{X \in_u S}[M_i(X^i) = 1] \leq (1 + \eta)^n \Pr_{X \in_u S}[\sum_j X_{ij} \leq r_i]. \tag{B.1}$$

Next we show how to efficiently compute the probability of all the $M_i$s accepting simultaneously under $S$.

**Lemma B.3.** *We can in time $(n^{O(m)}(\log R)/\eta)^m$ compute $\Pr_{X \in_u S}[\wedge_{i=1}^m M_i(X^i) = 1]$.*

We first prove Theorem 1.4 assuming these lemmas.

*Proof of Theorem 1.4.* Set $\eta = \varepsilon/mn^{m+1}$ in Lemma B.2 to obtain interval ROBPs $M_1, \ldots, M_m$ satisfying Equation B.1. Then, $X \in S$ implies $M_i(X^i) = 1$ for $i \in [m]$ and by a union bound,

$$\Pr_{X \in_u S}[\wedge_{i=1}^m M_i(X^i) \neq \mathbf{1}\{X \in CT'(r,c)\}] \leq \frac{\varepsilon}{n^m}.$$

On the other hand, by Lemma B.1,

$$\Pr_{X \in_u S}[X \in CT'(r,c)] \geq \frac{1}{n^m}.$$

Combining the above two equations we get that

$$\Pr_{X \in_u S}[X \in CT'(r,c)] \leq \Pr_{X \in_u S}[\wedge_{i=1}^m M_i(X^i) = 1] \leq (1 + \varepsilon) \Pr_{X \in_u S}[X \in CT'(r,c)].$$

Thus, $p = |S| \Pr_{X \in_u S}[\wedge_{i=1}^m M_i(X^i) = 1]$ is a $\varepsilon$-relative error approximation for $|CT'(r,c)| = |S| \Pr_{X \in_u S}[X \in CT'(r,c)]$. The theorem now follows as by Lemmas B.1, B.3, we can compute $p$ deterministically in time $(n^{O(m)}(\log R)/\varepsilon)^m$. □

## B.1 Proof of Lemma B.2

We show how to construct $M_1$; the constructions of $M_2, \ldots, M_m$ are similar. As in Section A we start with an interval ROBP $M$ that exactly computes the function $M(x) = 1\{\sum_j x_j \leq r_1\}$ and compute a sequence of interval ROBPs $M^n = M \leq M^{n-1} \leq \cdots \leq M^0$, where $M^i$ is obtained by *rounding* $M^{i+1}$. The final ROBP $M^0$ will have width at most $n^{O(m)}(\log R)/\eta$. Throughout this section, without explicitly saying so, we shall assume that all interval ROBPs are stored and computed succinctly as was done in Section A.

We now describe how to get $M^i$ from $M^{i+1}$. For $u \in T$, and $l \in [n]$, let $D(u, l)$ be the distribution of $(X_{l+1}, \ldots, X_n) \in \mathbb{Z}_+^{m \times (n-l)}$ for $X \in_u S$ conditioned on $\sum_{k \leq l} h(X_k) = u$. Further, let $D_1(u, l)$ denote the distribution of the first row of $Y$ for $Y \leftarrow D(u, l)$.

For a vertex $v$ in layer $l$ of $M^j$, $j \leq n$ and $u \in T$, let

$$P_{M^{i+1}, u}(v) = \Pr_{x \leftarrow D_1(u,l)}[\text{starting from } v, x \text{ leads to an accepting state in } M^{i+1}].$$

Let $L(M^{i+1}, i+1) = \{v_1 \prec v_2 \prec \cdots\}$. Fix $u \in T$ and define a set $B(u) = \{v_{u(j)}\} \subseteq L(M^{i+1}, i+1)$ of breakpoints for $u$ as follows. Start with $v_{u(1)} = v_R$ and given $v_{u(j)}$ define $v_{u(j+1)}$ by

$$v_{u(j+1)} = \max v \text{ such that } v \prec v_{u(j)} \text{ and } 0 < P_{M^{i+1},u}(v) < P_{M^{i+1},u}(v_{u(j)})/(1+\eta). \tag{B.2}$$

We set $L(M^i, i+1) = \cup_{u \in T} B(u)$ to be the union of the breakpoints for all $u$. Let $L(M^i, i+1) = \{b_1 \prec \cdots \prec b_N\}$. Note that $N < (2n^2)^m (n \log R)/\eta$. The vertices in all other layers stay the same as in $M^{i+1}$, as do all the edges except those from layer $i$ to $i+1$. We round these edge *upward* as before: let $v \in L(M^{i+1}, i)$ and for an edge label $b$, $M^{i+1}(v, b) = v' \in L(M^{i+1}, i+1)$. Find two consecutive vertices $b_k, b_{k+1} \in L(M^i, i+1)$ such that $b_k \prec v' \preceq b_{k+1}$. We set $M^i(v, b) = b_{k+1}$. Since the analysis of $M^0$ is similar to the analysis of Lemmas 4.5, 4.7 and A.3, we only analyze the complexity of constructing $M^0$ which is a little trickier. We need the following preliminary lemmas which are implicit in Dyer's FPRAS for contingency tables.

**Lemma B.4** (Implicit in Dyer). *For $j \in [n]$, and intervals $I_1, \ldots, I_m \subseteq [0, R]$, we can estimate $\Pr_{y \in_u T(j)}[\wedge_{k=1}^m \in (y_k \in I_k)]$ in time $O(m 2^m)$.*

*Proof.* Follows from an argument similar that of Lemma 4 in Dyer. □

**Lemma B.5** (Implicit in Dyer). *For $u, z \in T$ and $l \in [n]$, $(X_{l+1}, \ldots, X_n) \leftarrow D(u, l)$, we can estimate $\Pr[h(X_{l+1}) = z]$ in time $O((2n)^{4m+1})$.*

*Proof.* For $t \in T$, let $\delta_j(t) = |\{x \in T(j) : h(x) = t\}|$ and

$$f(k, t) = |\{(y_{k+1}, \ldots, y_n) \in T(k+1) \times \cdots \times T(n) : \sum_{i=k+1}^n h(y_i) \leq 2n^2 1 - t\}|.$$

Then, $\delta_j(t)$ can be computed in time $O(m 2^m)$ by the above lemma. Further, $f(n, t) = \delta_n(t)$ and for $k < n, s \in T$,

$$f(k, s) = \sum_{t \in T} \delta_{k+1}(t) f(k+1, t+s).$$

Therefore, we can compute $f(k, t)$ for all $k \in [n], t \in T$ in time $n^{4m+1}(m 2^m) = O((2n)^{4m+1})$. The lemma now follows as

$$\Pr[h(X_{l+1}) = z] = \frac{f(l+1, u+z)}{f(l, u)}.$$

□

We next show how to compute the transition probabilities $P_{M^i,u}(v)$ efficiently.

**Lemma B.6.** *For $i \in [n], u \in T$ and we can compute $L(M^i, i+1)$ and $P_{M^i,u}(v)$ for $v \in L(M^i, i+1)$ in time $n^{O(m)}(\log R)^3/\eta$.*

*Proof.* The proof is by induction on $n - i$. For $i = n$ there's nothing to show and suppose that for each vertex $w \in L(M^i, i+1)$ and $u \in T$, we know the values of $P_{M^i,u}(w)$. Fix a $v \in L(M^i, i)$, and let $X = (X_{i+1}, \ldots, X_n) \leftarrow D(u, i)$.

Now, from the definition of $D(u, i)$, given $h(X_{i+1}) = z$, $(X_{i+2}, \ldots, X_n)$ is independent of $X_{i+1}$ and moreover, the distribution of $(X_{i+2}, \ldots, X_n)$ is precisely $D(u + z, i + 2)$. Therefore,

$$P_{M^i,u}(v) = \sum_{w \in L(M^i, i+1)} \sum_{z \in T} \Pr_X[X_{1(i+1)} \in E(v, w) \wedge h(X_{i+1}) = z] \cdot P_{M^i,(u+z)}(w), \tag{B.3}$$

where $E(v, w) = \{y \in \mathbb{Z}_+ : M^i(v, y) = w\}$. Now, by the construction of $M^i$, $|L(M^i, i+1)| < (2n^2)^m (n \log R)/\eta = n^{O(m)}(\log R)/\eta$ and by the induction hypothesis, we know the values of $P_{M^i,(u+z)}(w)$ for all $w \in L(M^i, i+1)$. Thus to it is enough to show that we can compute $\Pr_x[x_{i+1} \in E(v, w) \wedge h_1(x_{i+1}) = z]$ efficiently for every $w$.

Fix a $w = w_j \in L(M^i, i+1)$. Then, $E(v, w_j) = \{y_1 \in \mathbb{Z}_+ : w_{j-1} \leq v + y_1 < w_j\}$ is an interval whose boundaries we know and let $I_1 = E(v, w_j) \cap \{y : h_1(y) = z_1\}$. Then, for $q = \Pr[X_{1(i+1)} \in E(v, w_j) \wedge h(X_{i+1}) = z]$,

$$q = \Pr[h(X_{i+1}) = z] \cdot \Pr[X_{1(i+1)} \in E(v, w_j) \mid h(X_{i+1}) = z]$$
$$= \Pr[h(X_{i+1}) = z] \cdot \Pr_{y \in_u T(i+1)}[y_1 \in E(v, w_j) \mid h(y) = z]$$
$$\text{(as conditioned on } h(X_{i+1}) = z, X_{i+1} \text{ is independent of } (X_{i+2}, \ldots, X_n)) \tag{B.4}$$
$$= \Pr[h(X_{i+1}) = z] \cdot \frac{\Pr_{y \in_u T(i+1)}[y_1 \in E(v, w_j) \wedge h(y) = z]}{\Pr_{y \in_u T(i+1)}[h(y) = z]}$$
$$= \Pr[h(X_{i+1}) = z] \cdot \frac{\Pr_{y \in_u T(i+1)}[y_1 \in I_1 \wedge_{k=2}^m h(y_k) = z_k]}{\Pr_{y \in_u T(i+1)}[h(y) = z]}. \tag{B.5}$$

Now, as $\{y_k : h(y_k) = z_k\} = \{y_k \in \mathbb{Z}_+ : r_k z_k/(2n^2) \leq y_k \leq r_k(z_k + 1)/(2n^2)\}$, it follows from Lemma B.4 that we can compute the second term above in time $n^{O(m)}$. Further, we can compute the first term efficiently by Lemma B.5. Thus, $q$ can be computed in time $n^{O(m)}$.

Thus, for each $v \in L(M^i, i), u \in T$, we can compute $P_{M^i,u}(v)$ in time $n^{O(m)}(\log R)/\eta$ and hence for a fixed $u \in T$, each new breakpoint in $B(u)$ can be found in time $n^{O(m)}(\log R)^2/\eta$ using binary search as in Lemma 3.2. Further, as $|L(M^{i-1}, i)| \leq n^{O(m)}(\log R)/\eta$ and we can compute $L(M^{i-1}, i)$ in time $n^{O(m)}(\log R)^3/\eta^2$ as claimed. $\square$

Lemma B.2 now follows from the above lemma and a straightforward extension of the arguments of Lemmas 4.4, 4.5, 4.7. We omit these details.

## B.2 Proof of Lemma B.3

The proof is similar to that of Lemma B.6: we show by induction that for $i \in [n], v_k \in L(M_k, i), 1 \leq k \leq m$ and $u \in T$ and $X \leftarrow D(u, i)$, we can compute

$$P(v_1, \ldots, v_m, u) = \text{Probability } X \text{ leads to an accepting when starting from } v_k \text{ in } M_k, \forall k \in [m].$$

For $i = n$, there's nothing to show and suppose the statement is true for $j \geq i+1$. Fix $v_1 \in L(M_k, 1), \ldots, v_m \in L(M_k, i)$ and $u \in T$. For $w_k \in L(M_k, i+1) \equiv L_k$, let $E_k(v_k, w_k) = \{y \in \mathbb{Z}_+ : M_k(v_k, y) = w_k\}$. Then, similar to Equation B.3, for $X = (X_{i+1}, \ldots, X_n) \leftarrow D(u, i)$,

$$P(v_1, \ldots, v_m, u) = \sum_{(w_1, \ldots, w_m): w_k \in L_k} \sum_{z \in T} \Pr[\wedge_k(X_{k(i+1)} \in E(v_k, w_k)) \wedge h(X_{i+1}) = z] \cdot P(w_1, \ldots, w_m, u+z). \tag{B.6}$$

We next show that for fixed $w_1 \in L_1, \ldots, w_m \in L_m, z \in T$, we can compute $q = \Pr[\wedge_k(X_{k(i+1)} \in E(v_k, w_k)) \wedge h(X_{i+1}) = z]$ in time $n^{O(m)}$. Let $I_k = \{y : y \in E(v_k, w_k)\} \cap \{y : h_k(y) = z_k\}$. Then, similar to Equation B.4, we have

$$\begin{aligned}
q &= \Pr[h(X_{i+1}) = z] \cdot \Pr[\wedge_k (X_{k(i+1)} \in E(v_k, w_k)) \,|\, h(X_{i+1}) = z] \\
&= \Pr[h(X_{i+1}) = z] \cdot \Pr_{y \in_u T(i+1)}[\wedge_k (y_k \in E(v_k, w_k)) \,|\, h(y) = z] \\
&= \Pr[h(X_{i+1}) = z] \cdot \frac{\Pr_{y \in_u T(i+1)}[\wedge_k (y_k \in I_k)]}{\Pr_{y \in_u T(i+1)}[h(y) = z]}.
\end{aligned}$$

Combining the above equation with Lemmas B.4, B.5, we can compute $q$ in time $n^{O(m)}$. Therefore, by Equation B.6, we can compute $P(v_1, \ldots, v_m, u)$ in time $(n^{O(m)}(\log R)/\eta)^m$. Lemma B.3 now follows by induction.